%% file: main.tex
\documentclass{PoS}

\input{preamble}

\title{Meson-meson scattering lengths at maximum isospin from lattice QCD}

\ShortTitle{Lattice QCD meson-meson scattering lengths}

\author{C.~Helmes,
C.~Jost,
B.~Knippschild,
\speaker{B.~Kostrzewa}\thanks{\texttt{kostrzewa@hiskp.uni-bonn.de}},
F.~Pittler,
C.~Urbach,
M.~Werner\\
Helmholtz Institut f{\"u}r Strahlen- und Kernphysik, University of Bonn, Bonn, Germany
} 

\author{L.~Liu\\
Institute of Modern Physics, Chinese Academy of Sciences, Lanzhou, China}

\abstract{We summarize our lattice QCD determinations of the pion-pion, pion-kaon and kaon-kaon s-wave scattering lengths at maximal isospin with a particular focus on the extrapolation to the physical point and the usage of next-to-leading order chiral perturbation theory to do so.
We employ data at three values of the lattice spacing and pion masses ranging from around 230 MeV to around 450 MeV, applying Luescher's finite volume method to compute the scattering lengths.
We find that leading order chiral perturbation theory is surprisingly close to our data even in the kaon-kaon case for our entire range of pion masses.}

\FullConference{The 9th International workshop on Chiral Dynamics\\
		17-21 September 2018\\
		Durham, NC, USA}

\begin{document}

\section{Introduction}

In this contribution we summarise our calculations of the $I=2, \pi\pi$~\cite{Helmes:2015gla}, $I=1, KK$~\cite{Helmes:2017smr} and $I=3/2, \pi K$~\cite{Helmes:2018nug} s-wave scattering lengths from lattice QCD.
We employ a set of gauge configuration ensembles generated by the European Twisted Mass Collaboration (ETMC) with $N_f=2+1+1$ quark flavours~\cite{Baron:2010bv} encompassing three lattice spacings and pion masses between $230$ and $450$ MeV, allowing a controlled extrapolation to the physical point and continuum limit.

\section{Lattice and finite volume methodology}

We use the twisted mass action, for which the Dirac operator of the light quark doublet reads~\cite{Frezzotti:2000nk} 
\begin{equation}
  D_\ell = D_\mathrm{W} + m_0 + i \mu_\ell \gamma_5\tau^3\, ,
  \label{eq:Dlight}
\end{equation}
where $D_\mathrm{W}$ denotes the standard Wilson Dirac operator and $\mu_\ell$ the bare light twisted mass parameter. 
Here and below, $\tau^i, i=1,2,3$ are the Pauli matrices acting in flavour space.
$D_\ell$ acts on a spinor $\chi_\ell = (u,d)^T$ and, hence, the $u$ ($d$) quark has twisted mass $+\mu_\ell$ ($-\mu_\ell$).

For the heavy doublet of charm and strange quarks~\cite{Frezzotti:2003xj}, the Dirac operator is given by
\begin{equation}
  D_\mathrm{h} = D_\mathrm{W} + m_0 + i \mu_\sigma \gamma_5\tau^1 + \mu_\delta \tau^3\,.
  \label{eq:Dsc}
\end{equation}

The bare Wilson quark mass $m_0$ has been tuned to its critical value $m_\mathrm{crit}$~\cite{Chiarappa:2006ae,Baron:2010bv}.
This guarantees automatic $\mathcal{O}\left(a\right)$-improvement \cite{Frezzotti:2003ni}, which is one of the main advantages of the Wilson twisted mass formulation of lattice QCD.

To avoid parity and flavour mixing between strange and charm quarks due to the splitting term in \cref{eq:Dsc}, we rely on a mixed-action approach for the strange quark by using the so-called Osterwalder-Seiler (OS) discretisation~\cite{Frezzotti:2004wz} with Dirac operator
\begin{equation}
  \label{eq:DOS}
  D_s^\pm = D_\mathrm{W} + m_0 \pm i \mu_s \gamma_5\,,
\end{equation}
and bare strange quark mass $\mu_s$.
It was shown in Ref.~\cite{Frezzotti:2004wz} that $\mathcal{O}(a)$-improvement remains intact when $m_0$ is set to the same value $m_\mathrm{crit}$ as used in the sea sector.
For each $\beta$-value, we choose a set of three bare strange quark masses $a\mu^{1,2,3}_s$ such as to bracket the physical strange quark mass indepenently of the light quark mass.

The lattice scale for the ensembles has been determined in Ref.~\cite{Carrasco:2014cwa} using $f_\pi$.
Also in Ref.~\cite{Carrasco:2014cwa} the non-singlet pseudoscalar renormalisation constant $Z_P$, the inverse of which is the quark mass renormalisation constant in the twisted-mass approach, has been determined for each lattice spacing.

We are interested in the limit of small scattering momenta for the three systems in question below inelastic threshold.
The scattering lengths $a_0^I$ for isospin $I$ can be related in the finite range expansion to the energy shift $\delta E$ by an expansion in $1/L$ following Ref.~\cite{Luscher:1986pf}
\begin{equation}
  \label{eq:luscher1}
  \delta E_X = - \frac{4 \pi a_0} {2 \mu_X L^3} \left(1 +c_1 \frac{a^I_0}{L}
  + c_2 \frac{(a^I_0)^2}{L^2} + c_3\frac{(a^I_0)^3}{L^3}\right) -
  \frac{8\pi^2 (a_0^I)^3}{2 \mu_X L^6}r^I_f +
  \mathcal{O}(L^{-7})\,,
\end{equation}
with coefficients~\cite{Luscher:1986pf,Beane:2007qr}
\begin{displaymath}
c_1 =  -2.837297\,,\quad c_2=6.375183\,,\quad c_3=-8.311951\,,
\end{displaymath}
and where $X \in \lbrace \pi\pi, \pi K, KK \rbrace$ and thus the $\mu_X$ correspond to the reduced masses of the respective two boson systems
\begin{align*}
  \mu_{\pi\pi} &= \frac{1}{2} M_{\pi}  & \mu_{\pi K} &= \frac{M_\pi M_K}{M_\pi + M_K} & \mu_{KK} &= \frac{1}{2} M_K \,,
\end{align*}
while $r^I_f$ are the effective range parameters.

\begin{figure}[t]
  \includegraphics[width=0.46\textwidth, page=3]{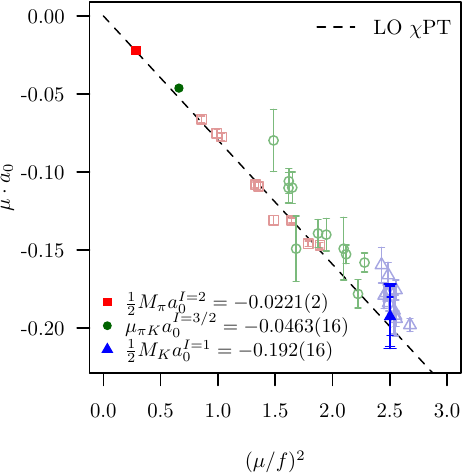}\hspace{0.3cm}
  \includegraphics[width=0.46\textwidth, page=4]{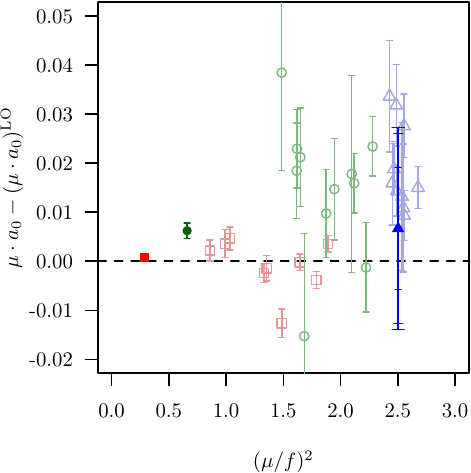}
  \caption{\textbf{(left)} Our data (open symbols) for all three systems in units as given in \cref{eq:mua0_lo}. The results of our continuum limit and extrapolation to the physical point are shown by the filled symbols while the dashed line indicates the LO $\chi$PT estimate. \textbf{(right)} Our data and final results with the LO $\chi$PT estimate subtracted.
  \label{fig:mua0_overview}}
\end{figure}

\section{Overview of our results}

Before detailing our three computations, we would like to point out that universal leading-order (LO) $\chi$PT is surprisingly close to our data for all three systems when written in the form
\begin{equation}
 \label{eq:mua0_lo}
 \mu_X \cdot a_0^I \underset{\mathrm{LO}}{=} - \frac{1}{4\pi} \left( \frac{\mu_X}{f_X} \right)^2 \,,
\end{equation}
where $\mu_X$ are as above and
\begin{align*}
  f_{\pi \pi} &= f_\pi & f_{\pi K} &= f_\pi  & f_{KK} &= f_K \,.
\end{align*}
It should be noted that one could also employ $f_\pi f_K$ for the $\pi K$ system which would bring the data even closer to the LO estimate.

\cref{fig:mua0_overview} gives an overview of all of our data for the scattering lengths indicated by the open symbols in the left panel.
The LO $\chi$PT estimate of \cref{eq:mua0_lo} is also shown by the dashed line, while the final results of our analyses in the continuum limit and at the physical point are given by the filled symbols.
The right panel instead shows the deviations of our data and final results from the LO estimate, which are seen to be statistically significant for the $\pi\pi$ and $\pi K$ cases but do not exceed a few percent even for our heaviest pion masses.

\section{$I=2, \pi\pi$}

\begin{figure}[t]
  \centering
  \includegraphics[width=0.46\textwidth,page=1]{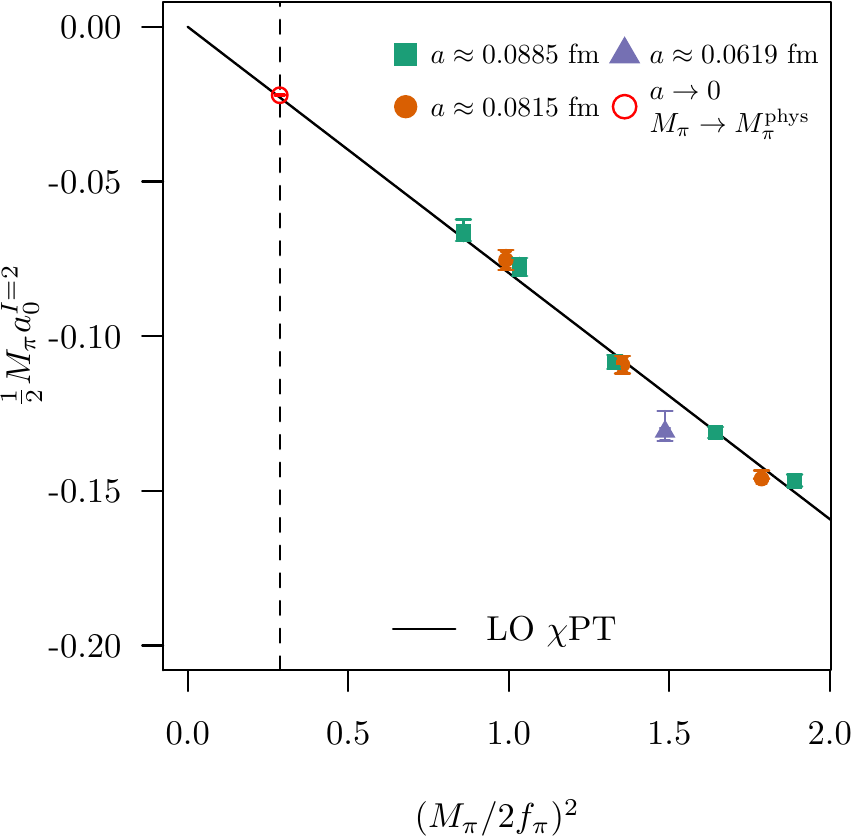}\hspace{0.3cm}
  \includegraphics[width=0.46\textwidth,page=2]{graphics/mpi_a0.pdf}
  \caption{\textbf{(left)} Overview of our data for $\mpiazero$ as a function of $(M_\pi/2f_\pi)^2$. The solid line indicates the LO $\chi$PT curve while the red empty circle indicates the result at the physical point in the continuum limit. \textbf{(right)} Difference between our results for $\mpiazero$ and the LO $\chi$PT estimate. The lines show the NLO piece from fits of \cref{eq:mpia0_nlo} with different $M_\pi/f_\pi$ cuts. The dark and light grey bands indicate the statistical and combined statistical and systematic errors, respectively, as given in \cref{eq:mpia0_final}.
  \label{fig:mpi_a0}}
\end{figure}

In order to extrapolate our data for $M_\pi a_0$ to the physical point, we employ next-to-leading order (NLO) continuum $\chi$PT.
As suggested in Refs.~\cite{Beane:2005rj,Beane:2007uh}, it is convenient to write the expression for $M_\pi a_0$ as a  function of $M_\pi/f_\pi$ because then all quantities are dimensionless and no scale input is needed.
This results in~\cite{Beane:2005rj,Beane:2007uh} 
\begin{equation}
  \label{eq:mpia0_nlo}
  M_\pi a_0 = -\frac{M_\pi^2}{8\pi f_\pi^2}\left\{1 +
    \frac{M_\pi^2}{16\pi^2 f_\pi^2} \left[3\ln \frac{M_\pi^2}{f_\pi^2}
    - 1 - \ell_{\pi\pi}(\Lambda_\chi = f_{\pi,\mathrm{phys}})\right]\right\}
\end{equation}
with $\ell_{\pi\pi}$ related to the Gasser-Leutwyler coefficients $\bar\ell_i$ as follows~\cite{Bijnens:1997vq}
\begin{displaymath}
\ell_{\pi\pi}(\Lambda_\chi) = \frac{8}{3} \bar\ell_1 +
\frac{16}{3}\bar\ell_2 -\bar\ell_3 -4\bar\ell_4 +
3\ln\frac{M_{\pi,\mathrm{phys}}^2}{\Lambda_\chi^2}\,.
\end{displaymath}
One can show in Wilson twisted mass $\chi$PT that the leading lattice artefacts to $M_\pi a_0$ are of $\mathcal{O}(a^2M_\pi^2)$~\cite{Buchoff:2008hh}, such that at NLO, we consistently describe our data with the continuum $\chi$PT formula provided above.

In the expression here and those in the subsequent sections, we formally fix the scale-dependent LECs at $\Lambda_\chi = f_{\pi^-}^\mathrm{phys}$.
In practice, however, it has proven useful to employ the values of the pion decay constant as measured on the lattice for each ensemble with finite size corrections applied.
Doing so has the benefit of giving statistically precise ratios and not requiring scale setting at the cost of only inducing higher order corrections in the chiral expansion.

We show our data for $\mpiazero$ together the LO $\chi$PT estimate in the left panel of \cref{fig:mpi_a0}.
In the right panel, instead, we have subtracted the LO estimate to clarify the size of corrections beyond the LO.
The NLO pieces from fits of \cref{eq:mpia0_nlo} with different cuts in $M_\pi/f_\pi$, as indicated by the brackets, are shown by the solid purple, blue and black lines.
Comparing these different fits, we obtain as our final result
\begin{equation}
M_\pi a_0\ =\
-0.0442(2)_\mathrm{stat}(^{+4}_{-0})_\mathrm{sys}\,,\qquad
\ell_{\pi\pi}\ =\
3.79(0.61)_\mathrm{stat}(^{+1.34}_{-0.11})_\mathrm{sys}\,, 
\label{eq:mpia0_final}
\end{equation}
where the first error is statistical only and the second accounts for the different cuts.
It turns out that other systematic errors are much smaller than the statistical uncertainty such that we can safely ignore them.
We would like to point out that after the LO contribution has been subtracted, it becomes clear that even though our data is rather precise compared to other lattice determinations, we are only just sensitive enough to determine the NLO corrections and that it is the constraint in the chiral limit which allows us to quote the very small error on our final result.

\begin{figure}
  \centering
  \includegraphics[width=0.8\textwidth]{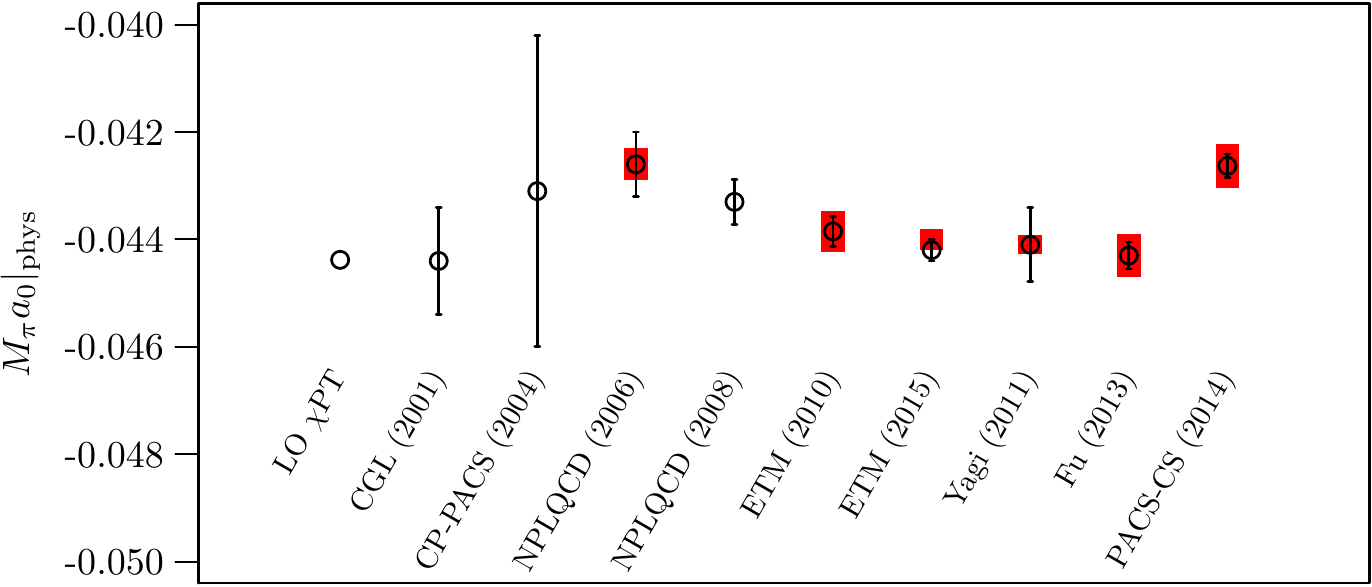}
  \caption{Comparison of this work (ETM 2015) to other determinations, for references see the main text.
  \label{fig:mpi_a0_compare}}
\end{figure}

In \cref{fig:mpi_a0_compare}, we compare our result to those of CP-PACS~\cite{Yamazaki:2004qb}, NPLQCD (2006)~\cite{Beane:2005rj}, NPLQCD (2008)~\cite{Beane:2007xs}, ETM (2013)~\cite{Feng:2009ij}, this work denoted as ETM (2015), Yagi et al.~\cite{Yagi:2011jn},
Fu~\cite{Fu:2013ffa} and PACS-CS~\cite{Sasaki:2013vxa}. 
We quote statistical and -- where available -- systematic uncertainties separately.
For NPLQCD~(2008) there is only the combined statistical and systematic uncertainty.

\section{$I=1, KK$}

For the $KK$ and $\pi K$ systems, we interpolate all of our data to fixed (physical) renormalised strange quark mass either by requiring $M_K^2 - \frac{1}{2} M_\pi^2$ to take its physical value (henceforth method $A$) for all of our ensembles or by defining per-ensemble reference strange quark masses taking into account lattice artefacts, such that $M_K^2$ parametrised in terms of SU(2) $\chi$PT at fixed strange quark mass takes its physical value in the continuum limit extrapolated to the physical point (method $B$).
For details, see Refs.\cite{Helmes:2017smr,Helmes:2018nug}.

\begin{figure}
  \centering
  \includegraphics[width=0.46\textwidth, page=3]{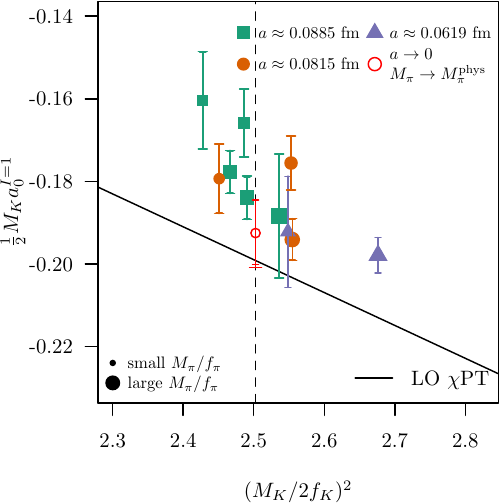}\hspace{0.3cm}
  \includegraphics[width=0.46\textwidth, page=4]{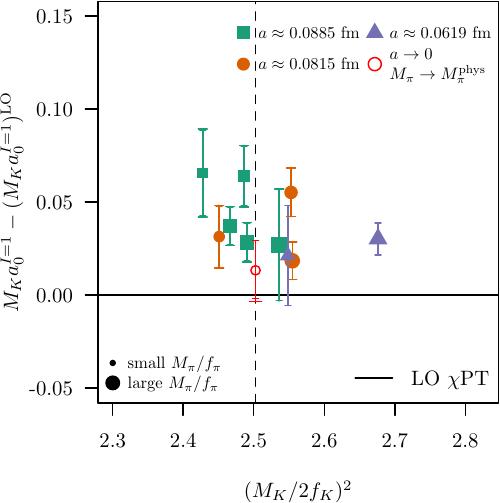}
  \caption{\textbf{(left)} Overview of our data for $\mkazero$ as a function of the reduced mass of the $\pi K$ system squared in units of the pion decay constant. The solid line indicates the LO $\chi$PT curve while the red empty circle indicates the result at the physical point in the continuum limit. \textbf{(right)} Deviation of our data and final result from the LO $\chi$PT estimate.
  \label{fig:mk_a0}}
\end{figure}

Working at fixed strange quark mass, we attempt to fit the NLO expression for the $I=1, KK$ scattering length, given in Refs.~\cite{Gasser:1984ux,Bernard:1990kw,Chen:2006wf} as
\begin{equation}
  \label{eq:mka0_nlo}
  M_K a_0 = - \frac{M_K^2}{8\pi f_K^2}\left[1 - \frac{16}{f_K^2}\left(
    M_K^2 L^\prime -\frac{M_K^2}{2} L_5 + \zeta\right) \right]\,,
\end{equation}
where $L_5$ is a low energy constant, $L'$ is a combination of standard low energy constants and $\zeta$ is a known function of meson masses and chiral logarithms.
Using the normalisation of \cref{eq:mua0_lo}, our data together with the LO $\chi$PT prediction is shown in the left panel of \cref{fig:mk_a0}.
In the right panel, instead, we subtract the LO prediction and also here it is seen that contributions beyond the LO are genuinely small.

As our strange quark mass is tuned very close to its physical value, our data does not posess sufficient spread in $M_K^2/f_K^2$ to allow for an independent determination of $L_5$ and $L'$ together with a check of residual discretisation effects, which appear to be marginally visible.
Using valence strange quark masses far from the physical value to increase the spread in $M_K^2/f_K^2$ instead would potentially lead to unitarity breaking contributions which cannot be neglected, such that we abstained from this approach.

\begin{figure}
  \centering
  \includegraphics[width=0.46\textwidth]{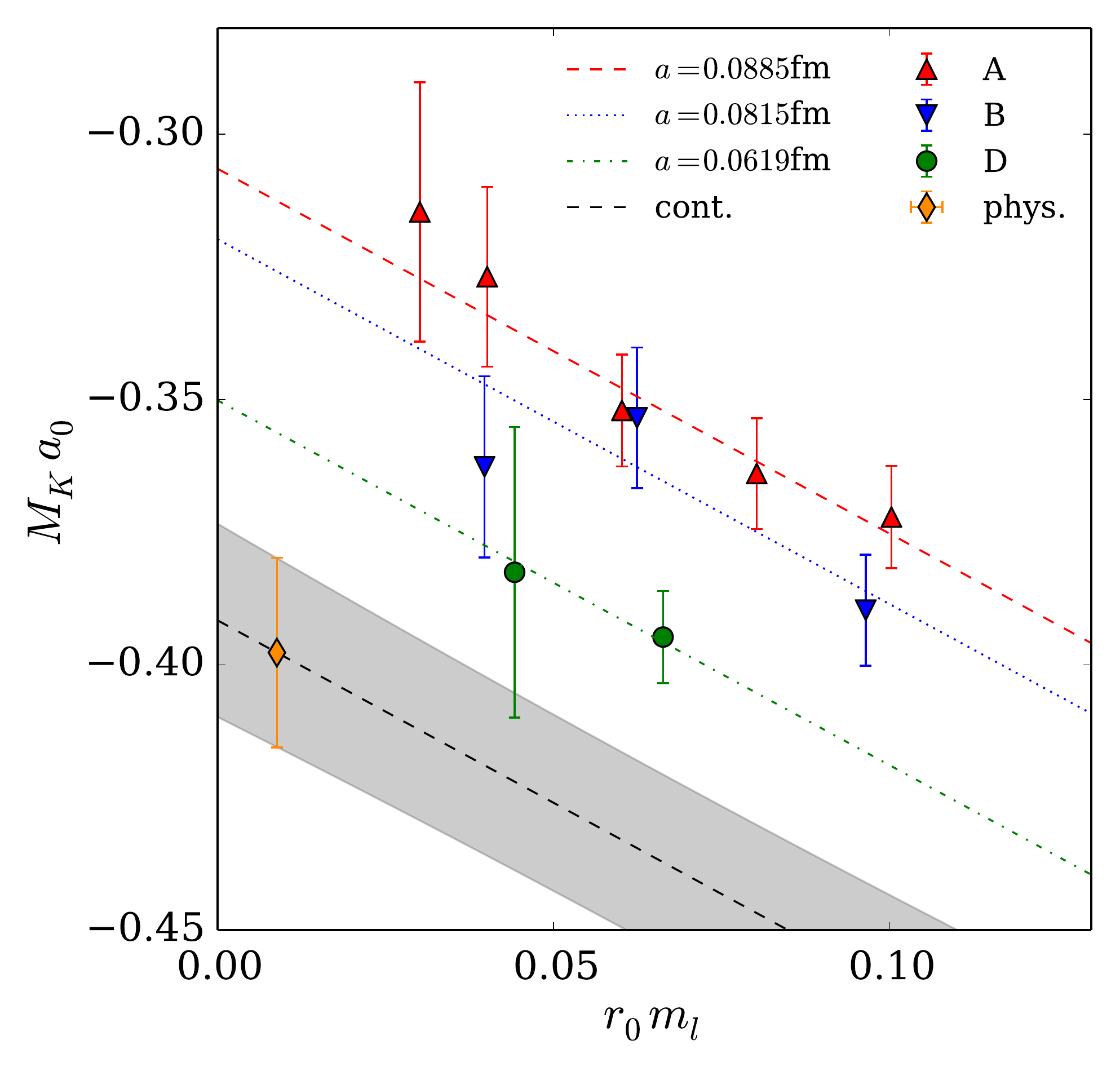}
  \includegraphics[width=0.46\textwidth]{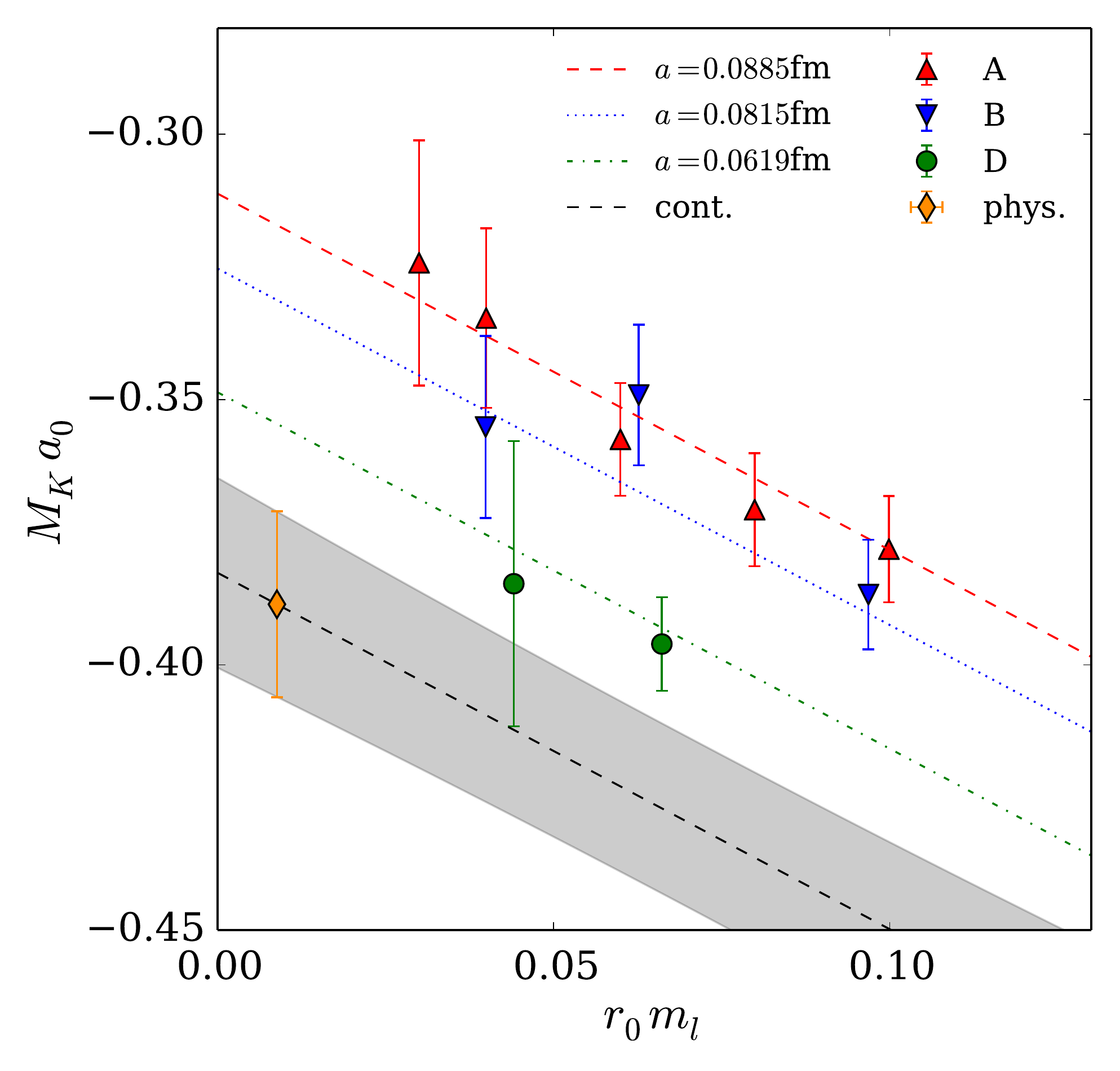}
  \caption{Chiral and continuum extrapolation of $\mkazero$ as a function of the renormalised light quark mass in units of the Sommer scale $r_0$~\cite{Sommer:1993ce} at fixed (physical) strange quark mass. \textbf{(left)} and \textbf{(right)} differ in how the strange quark mass is fixed to its physical value as described above \cref{eq:mka0_nlo}.
  \label{fig:MK_a0_at_mk_ms}}
\end{figure}

Instead, we perform an explicit extrapolation to the continuum limit and the physical value of the renormalised light quark mass in a global fit with the Ansatz
\begin{equation}
  \label{eq:mka0_chi_cont_ext}
  M_K a_0 = Q_0 \frac{P_r}{P_Z} a\mu_\ell + Q_1 \frac{1}{P_r^2} + Q_2\,,
\end{equation}
where $a\mu_\ell$ is the bare light quark mass in lattice units, $Q_i$ are simple fit parameters while $P_r$ and $P_z$ are restricted by priors for the Sommer scale $r_0$~\cite{Sommer:1993ce} and the non-singlet pseudoscalar renormlisation constant $Z_P$, respectively.
$Q_1$ accounts for lattice artefacts which are clearly visible when the data is parametrised in terms of the light quark mass, as shown in \cref{fig:MK_a0_at_mk_ms}.
There, the left panel displays the data interpolated to the value of the strange quark mass fixed via method $A$ and the right panel when method $B$ is used instead.
Our final result reads
\begin{equation}
M_K a_0 = -0.385(16)_{\textrm{stat}} (^{+0}_{-12})_{m_s}(^{+0}_{-5})_{Z_P}(4)_{r_f} \,,
\label{eq:mka0_final}
\end{equation}
where the first error is statistical only, the second accounts for the two ways of interpolating the data to the physical strange quark mass, the third accounts for two different determinations of the renormalisation constant while the last one indicates the effect of neglecting higher order terms in the calculation of the scattering length via the L{\"u}scher method.

\begin{SCfigure}[0.65]
  \includegraphics[width=0.65\textwidth]{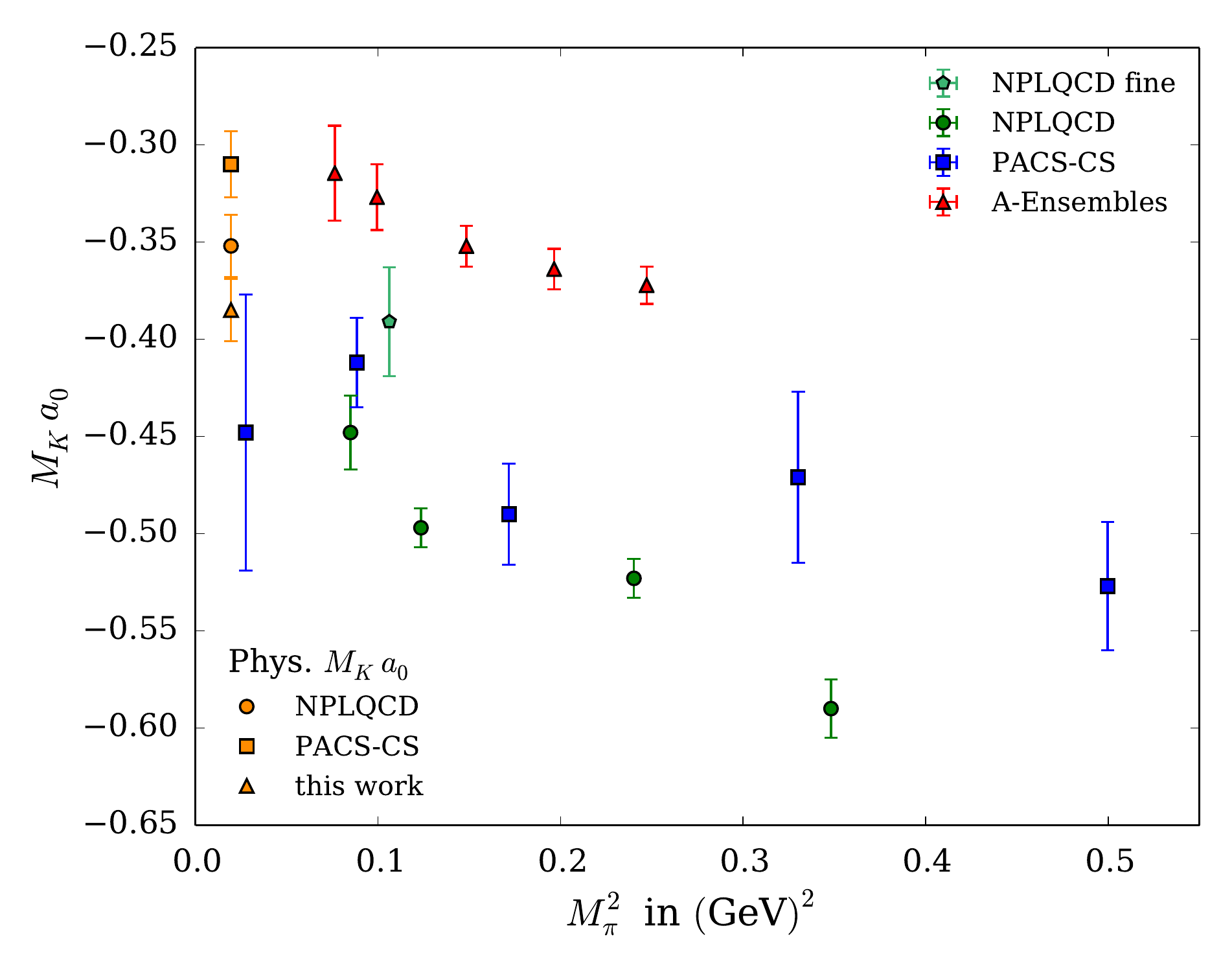}
  \caption{$\mkazero$ as a function of $M_\pi^2$. The orange triangle shows our final result while the red triangles show our data at the coarsest lattice spacing for one method of fixing the strange quark mass and determination of the quark mass renormalisation constant.
  The final results of NPLQCD~\cite{Beane:2007uh} and PACS-CS~\cite{Sasaki:2013vxa} are indicated by the orange circle and square, respectively.
  \label{fig:mka0_comp}}
\end{SCfigure}

We can compare this this to the only other two lattice computations of Refs.~\cite{Beane:2007uh,Sasaki:2013vxa} in \cref{fig:mka0_comp} where it becomes clear that lattice artefacts should not be neglected, although we can confirm that when parametrsied via the continuum $\chi$PT form of \cref{eq:mka0_nlo}, substantial cancellations of lattice artefacts seem to occur, as claimed in Ref.\cite{Chen:2006wf}.

\section{$I=3/2, \pi K$ scattering}

The NLO $\chi$PT expression for $\mupikazero$ can be derived from Refs.~\cite{Bernard:1990kw,Kubis:2001ij} giving
\begin{equation}
\resizebox{\textwidth}{!}{
  $\mupikazero = - \frac{\mu^2_{\pi K}}{4\pi f_\pi^2}
    \left[1 - \frac{32M_\pi M_K}{f_\pi^2} L_{\pi K}(\Lambda_\chi) +
      \frac{16 M_\pi^2}{f_\pi^2}L_{5}(\Lambda_\chi)
     -\frac{1}{16\pi^2 f_\pi^2} \chi_{\text{NLO}}^{3/2}(\Lambda_\chi,M_\pi,M_K,M_\eta)\right] 
     +c\cdot f(a^2)\,,$
}
\label{eq:mupika0_nlo}
\end{equation}
where we add the term $c\cdot f(a^2)$ to account for possible lattice artefacts as, unlike in the $\pi\pi$ and $KK$ cases, these might enter at NLO.
$L_5$ is the same as in \cref{eq:mka0_nlo} while $L_{\pi K}$ is a different combination of LECs and $\chi^{3/2}_{\mathrm{NLO}}$ is a function of meson masses and chiral logarithms.

\begin{figure}
  \centering
  \includegraphics[width=0.46\textwidth, page=1]{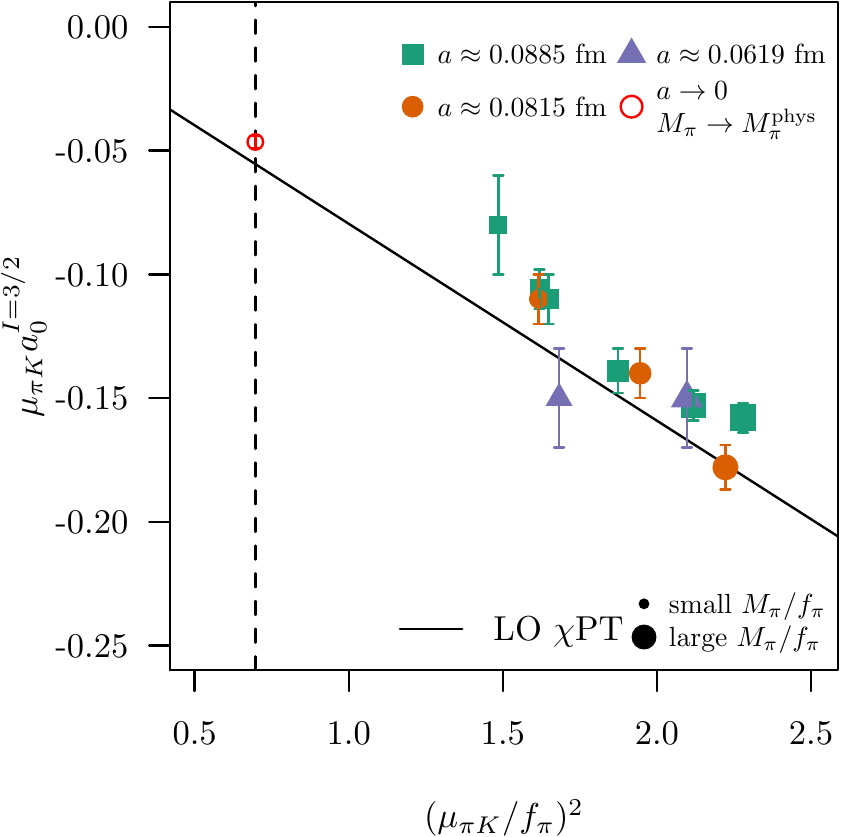}\hspace{0.3cm}
  \includegraphics[width=0.46\textwidth, page=2]{graphics/mupik_a0.pdf}
  \caption{\textbf{(left)} Overview of our data for $\mupikazero$ as a function of the reduced mass of the $\pi K$ system squared in units of the pion decay constant squared. The solid line indicates the LO $\chi$PT curve while the red empty circle indicates the result at the physical point in the continuum limit. \textbf{(right)} Our data and final result with the LO $\chi$PT estimate subtracted.
  \label{fig:mupik_a0}}
\end{figure}

In \cref{fig:mupik_a0} we show our data for $\mupikazero$ together with the LO $\chi$PT prediction in the left panel, while in the right panel we again subtract this from our data and final result, the latter of which is shown by the empty red circle.
Again it is clear that the deviation from LO is small, although here it seems to be consistent across our set of ensembles.

Unlike in the $KK$ case, we have a sufficient lever to determine $L_{\pi K}$, although we are unable to determine $L_5$ at the same time.
We constrain the latter with a prior based on the value determined in Ref.~\cite{Dowdall:2013rya} and translated to our renormalisation scale.
We are also unable to get a statistically significant result for fit parameter $c$ employing different kinds of functions to describe potential lattice artefacts, suggesting that they are negligible within our statistical uncertainties.

\begin{SCfigure}[0.6]
  \includegraphics[width=0.55\textwidth]{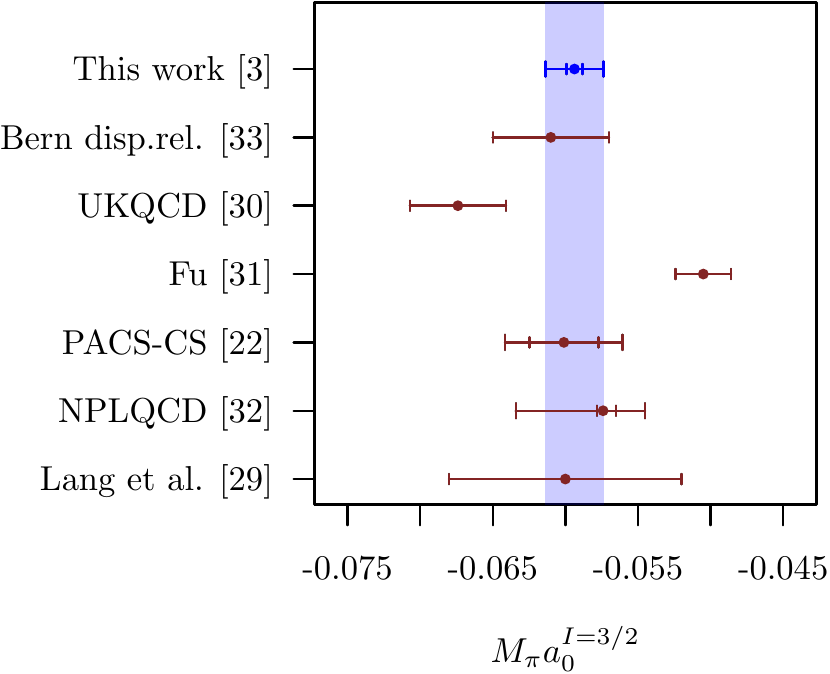}
  \caption{Comparison between our~\cite{Helmes:2018nug} determination of the $I=3/2, \pi K$ scattering length, various lattice calculations~\cite{Lang:2012sv,Janowski:2014uda,Fu:2011wc,Sasaki:2013vxa,Beane:2006gj} as well as a determination from ChPT presented at this conference ~\cite{ELVIRA}.
  \label{fig:mpi_pik_a0_compare}}
\end{SCfigure}

Our final result reads
\begin{equation}
\begin{split}
 \mpiathreehalveszero & = -0.05937 (82)_\mathrm{stat} (139)_\mathrm{fr} (64)_\mathrm{tp} (9)_{\chi\mathrm{PT}} \\
 L_{\pi K} & = 0.00379 (15)_\mathrm{stat} (19)_\mathrm{fr} (12)_\mathrm{tp} (1)_{\chi\mathrm{PT}} \,,
\end{split}
\end{equation}
where the first error is purely statistical and the second is an estimate of the systematic error due to fit range choice in the determination of the interacting energy.
The third is the systematic error stemming from two different ways to account for intermediate state contributions to the interacting energy, see Ref.~\cite{Helmes:2018nug} for details.
Finally, the last error gives the difference between fitting \cref{eq:mupika0_nlo} directly and employing the so-called Gamma-method~\cite{Beane:2006gj}.

We close this section by comparing our determination of $a_0^{I=3/2}$ in units of $M_\pi$ and the results of \cite{Lang:2012sv,Janowski:2014uda,Fu:2011wc,Sasaki:2013vxa,Beane:2006gj,ELVIRA} in \cref{fig:mpi_pik_a0_compare}.
The preliminary result of \cite{ELVIRA} has also been presented at this conference and we thank the authors for sharing the value and an estimate of the uncertainty.
The comparison shows that five out of seven determinations are fully compatible with each other, while the results of Refs.~\cite{Fu:2011wc,Janowski:2014uda} are somewhat high and low respectively.
This may be explained by lattice artefacts as both studies employ a single lattice spacing, although the latter is at the physical pion mass.

\section{Conlusions and Outlook}

We have presented determinations of the s-wave scattering lengths at maximal isospin for the $\pi\pi$, $KK$ and $\pi K$ systems from lattice QCD, for the first time employing multiple lattice spacings and a wide range of pion masses throughout.
We confirm that lattice artefacts are strongly suppressed when continuum $\chi$PT is used to extrapolate to the physical point.
At the same time, however, it is clear that the corrections beyond LO are very small and that even higher statistical precision is necessary to really constrain the values of all LECs.
As an outlook, it would be interesting to refine our analysis procedure by performing a global fit of all three systems, perhaps even combined with an analysis of meson masses and decay constants.
A further refinement would be obtained with the availability of configurations directly at the physical point.

\vspace{-0.5cm}
\section*{Acknowledgements}
\vspace{-0.25cm}

This work was funded by the DFG as a project in the Sino-German CRC110.
The authors gratefully acknowledge the Gauss Centre for Supercomputing e.V. (www.gauss-centre.eu) for funding this project by providing computing time on the GCS Supercomputer JUQUEEN~\cite{juqueen} and the John von Neumann Institute for Computing (NIC) for computing time provided on the supercomputer JURECA~\cite{jureca} at J{\"u}lich Supercomputing Centre (JSC).

\bibliographystyle{3authors_notitle_links}
\bibliography{bibliography}{}

\end{document}

%% file: preamble.tex
\usepackage{todonotes}

\usepackage{amsmath}

\usepackage[capitalise]{cleveref}

\usepackage{sidecap}
\sidecaptionvpos{figure}{t}
  
\newcommand{\mkazero}{M_K a_0^{I=1}}
\newcommand{\mpiazero}{M_\pi a_0^{I=2}}
\newcommand{\mupikazero}{{\mu_{\pi K}} a_0^{I=3/2}}
\newcommand{\mpiathreehalveszero}{M_\pi a_0^{I=3/2}}

\let\OLDthebibliography\thebibliography
\renewcommand\thebibliography[1]{
  \OLDthebibliography{#1}
  \setlength{\parskip}{0pt}
  \setlength{\itemsep}{0pt plus 0.3ex}
}